%% file: make_astro.tex
\documentclass{mpe_report}

\usepackage{psfig,graphicx,epsfig}
\usepackage{color}
\usepackage{amsmath,amssymb,epic,eepic,array}

\begin{document}

\pagenumbering{arabic}
\setcounter{page}{226}

 \renewcommand{\FirstPageOfPaper }{226}\renewcommand{\LastPageOfPaper }{229}\include{./mpe_report_calderone}      \clearpage

\end{document}

%% file: mpe_report_calderone.tex
\title{MCS, a new approach to data treatment in astronomical projects}
    \author{G. Calderone\inst{1} \and L. Nicastro\inst{2}}
    \institute{Istituto Nazionale di Astrofisica, IASF,
    Via U. La Malfa 153, 90146 Palermo, Italy \and
    Istituto Nazionale di Astrofisica, IASF,
    Via P. Gobetti 101, 40129 Bologna, Italy}
 \maketitle

\begin{abstract}
Today's astronomical projects need computational systems capable to
store and analyze large amounts of scientific data, to effectively share
data with other research Institutes and to easily implement information
services to present data for different purposes (scientific,
maintenance, outreach, etc.). Due to the wide scenario of
astronomical projects there isn't yet a standardized approach to
implement the software needed to support all the requirements
of a project. The new
approach we propose here is the use of a unified model where all data
are stored into the same data base becoming available in different
forms, to different users with different privileges.
\end{abstract}

\section{Introduction}
Information services can be separated in two classes: those in which
the information produced are addressed to humans, and those in which
they are meant for the use by other software applications. In the
former case there is a quite standardized way to develop such an
information service, essentially based upon a web server, a database
server, a scripting language and HTML pages. In the latter case
instead there is no such standardization, and here's why
\textbf{MCS} (My Customizable Server) was implemented.

MCS is a set of software tools aimed at easily implement information
services, that is an application that provides a service over the
network. At the core of a MCS-based system there is a TCP server which
listen for user connections, and once a user is connected it will send
all the requested information. However the transmitted data aren't in
free format (like in a web page), but are packed in a well defined
fashion (using the MCS protocol) so that on the other side a software
can understand what is being sent. The MCS high level classes will
hide all code implementations related to multi-threading, networking,
database access, etc., and require no low-level knowledge of these
issues by the users. MCS can also be customized through the derivation
of some classes.
So MCS and its protocol are for software applications what a web
server and HTTP are for the WWW: a simple way to access data. In this
comparison customizing the MCS server is like writing a web page.

MCS is the evolution of a software named SDAMS (SPOrt Data Archiving
and Management System) \cite{sdams}, that was implemented to support
the SPOrt experiment (\cite{sport}), an Italian Space Agency (ASI)
funded project which has
been ``frozen" because of the problems with the Columbus module on the
International Space Station. At that time the approach used to build
SDAMS seemed to be easily portable to other experiments so its
features and usability has been generalized until it became the actual
MCS. In fact it is now used to manage the data collected by the
optical camera ROSS (\cite{ross}) mounted on the robotic telescope REM
(\cite{rem}) at La Silla, Chile (see Nicastro \& Calderone, this
volume).

MCS has been developed on the GNU/Linux platform and is released
under the GPL license. It can be freely downloaded from the site
\textsf{ross.iasfbo.inaf.it/mcs/}. This site contains all news,
updates, documentation and downloadable software packages.
The documentation is present as a user manual (in \textsf{pdf} format)
which focuses on the main aspects of an MCS-based application, describes
the server environment, commands, and the interfaces for various
languages, and as a technical reference documentation for all MCS's
classes (in html format, automatically generated using
\textsf{Doxygen}). The site is still under development, so check for
updates.

\section{The MCS architecture}
The MCS core is basically a set of high level C++ classes aimed at
providing all the functionalities of an application server, that is an
application that will listen for user's request coming from the
network, eventually will access a database and/or execute an external
program, then it will send the answer to the user using the MCS
protocol. The main features of application servers built with MCS are:

\begin{itemize}
\item easy to configure;
\item authentication and grant support;
\item secure connections (through SSL);
\item database access (MySQL);
\item base commands set, support to create new customized commands;
\item accessibility (as clients) from other languages such as C,
Fortran, IDL, PHP, etc.;
\item logging facility, etc.
\end{itemize}
All these features are already available, without performing any
customization. So to implement a simple service with the above
features you'll only need to install MCS and configure it through a
simple configuration file. 
The only code needed is as follows:
%
\begin{verbatim}
//Mandatory includes for every MCS-based appl.
#include <mcs.hh>
using namespace mcs;
//Main program
int main(int argc, char *argv[]) {
  //Start the server...
  Env* env = mcsStart("simplest");
  //...and wait for its end
  mcsWait(env);
}
\end{verbatim}
For more complex services you can customize the server behavior in several ways:

\begin{itemize}
\item adding external programs, either real external applications or
  batch lists of MCS commands;
\item adding SQL programs, to be executed on the database server;
\item adding customized commands, deriving the \verb|UserThread|
  class;
\item modifying the behavior of the server side thread, deriving the
  \verb|LocalThread| class;
\end{itemize}
\begin{figure}[t]
\centerline{\psfig{file=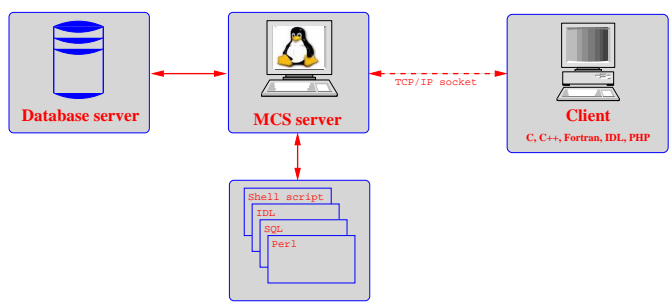,width=8.8cm,clip=} }
\caption{Main components of an MCS based system.}
\label{diaggen}
\end{figure}

In a typical application you should implement a database with all
the tables needed to store all the relevant data related
to the project, and eventually prepare the required external programs.
Figure \ref{diaggen} shows the typical architecture of an MCS-based system:
\\
\textbf{Database server}: the database server is used to handle
clients authentication, to store all application specific data and
anything else necessary to the application itself. This server isn't
accessible directly from the clients, but it is visible only to the
application server. At the moment the only supported database server
is MySQL\footnote{\textsf{www.mysql.com}}. In the future other servers may
become accessible through MCS.
\\
\textbf{Application server}: the application server is the core
of the information system. It implements the client/server model: a
client opens a TCP socket towards the host running MCS and sends a
request, then the server ``computes'' an answer, eventually querying
the database and/or executing some external programs, and sends it
back to the client.  The behavior of the MCS server can be customized
deriving some classes.
\\
\textbf{External programs}: external programs are software
applications written in any language, which interact with the
application server via command line and the standard output. Support to
these programs was added to easily integrate already existing
applications within MCS.
\\
\textbf{Clients}: clients are programs which access the MCS
service over the network. Such programs can be written in any language
and run on any platform, provided that they implement the MCS
protocol. Interfaces that implement the MCS protocol are provided by
the MCS library for the following languages on the Linux platform:
C++, C, Fortran, IDL, PHP. Support for other languages (such as Java,
Python and Perl) and the Windows platform will be available soon.

%
\begin{table}[t] \begin{center}
    \caption{MCS and Unix ``shell'' comparison.}\label{paragoneshell}
    \begin{tabular}{|l|l|}
      \hline \textbf{Unix shell} & \textbf{MCS server} \\
      \hline
      \verb|stdin| and \verb|stdout|    & bidirectional TCP socket \\
      system account                    & MySQL account \\
      internal commands                 & base commands \\ 
      programs, shell scripts           & external programs (\verb|EXEC| command) \\ 
      home directory                    & work directory \\
      \hline
    \end{tabular}
\end{center} \end{table}
Note that such a system would not be a closed model, in the sense that
existing databases and/or software tools may be integrated into the
MCS-based systems. That's because MCS doesn't make any assumption
about type and quantity of data you need to deal with. It uses a
relational database system to store many of its data.
Databases have become today the most useful, scalable and
flexible way to store and give access to any sort of data.

From a user's point of view MCS is very similar to the usage of a
classic Unix shell, that is a command line interface with a prompt on
which users can execute commands in their own environment and wait for
the output before a new command is issued. It is therefore possible to
make a comparison between the ``components'' of a shell, and the ones
from an MCS connection (see Tab. \ref{paragoneshell}).

\begin{figure}[hbtp]
\centerline{\psfig{file=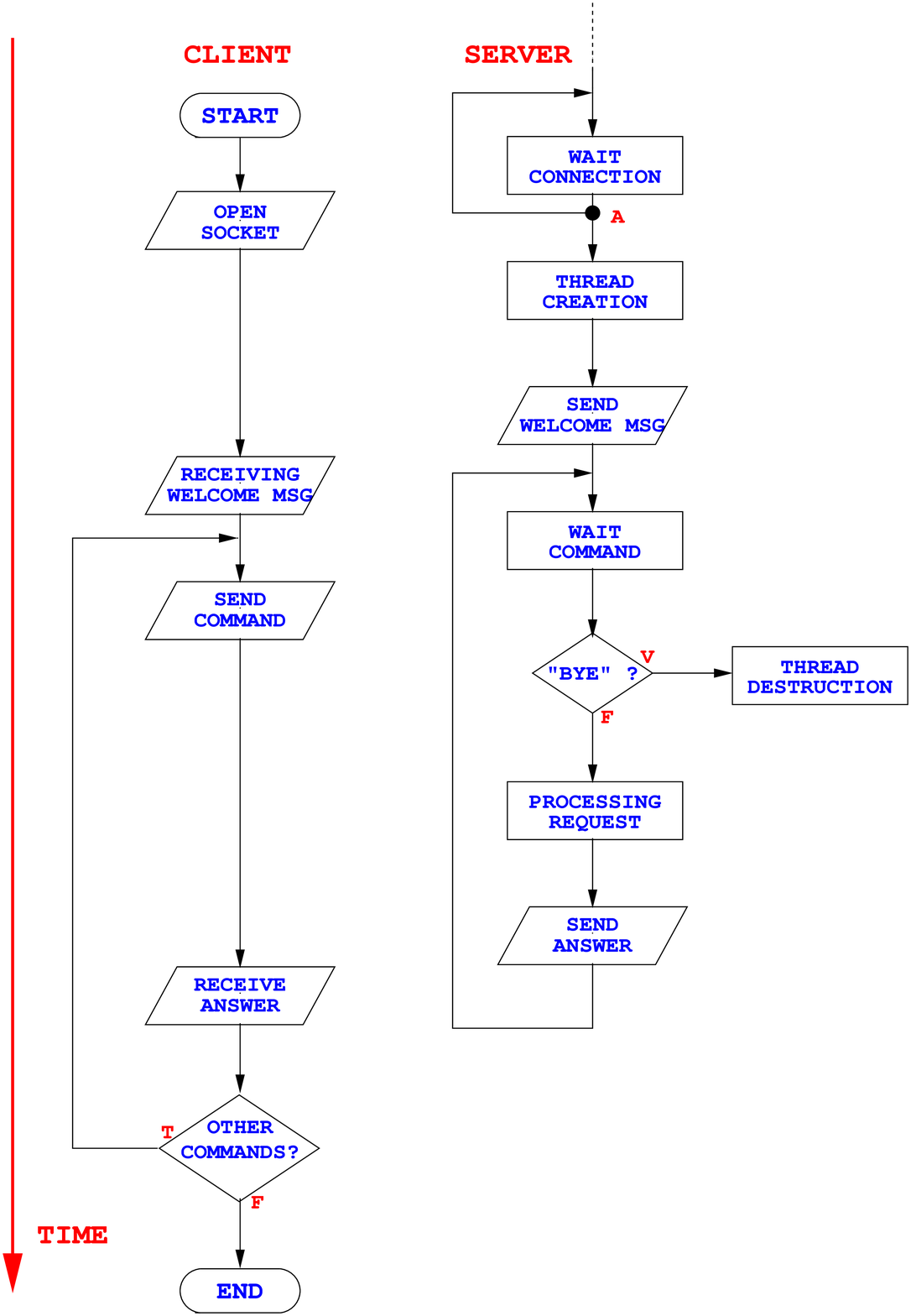,width=8.8cm,clip=} }
\caption{Flow diagram of a typical MCS session.}
\label{flow}
\end{figure}

The output of a command won't be ASCII text like in a shell.
Data will be instead formatted using the MCS protocol and can be sent
and received in binary form in both directions. Figure \ref{flow} shows
the typical sequence of events during an MCS session.

MCS classes can also be used outside an application server,
for example it is possible to write a program in C++ which uses
the \textbf{DBConn} and \textbf{Query} classes to access a database, or
one of the \textbf{VOT\_Parser\_Tree} or \textbf{VOT\_Parser\_Stream}
classes to read VOTable file. The same program can also be written in
one of the languages for which we have an interface (at the moment C,
Fortran, IDL, PHP; other will be developed in the near future).

\subsection{Deriving MCS classes}
Deriving a C++ class means creating a new class that has all the
behaviour characteristics of an existing one (the parent class), plus
some more specific behaviour added. In the MCS case the server
behaviour can be customized through the derivation of the
\verb|UserThread| class (more specifically only the ``virtual''
methods should be derived, as described in the documentation). This
way it is possible to create custom commands available as if they were
``base commands'' (see Tab. \ref{paragoneshell}). Another class
that can be derived for customization is \verb|LocalThread|. It runs
in a server side thread, independently from other client threads. This
class can be used to implement some server side tasks like data
quick look or reduction, database maintenance, etc.

\section{A real life example}
In a typical scientific experiment we have an instrument producing
data, a main storage system, a set of software tools to perform
analysis, and people with different needs who wish to access the
data. In this section we'll analyze the components of an informative
system based on MCS, applied to such an experiment.  A common use of
MCS in such a real life project would be as follows: an MCS server is
running on the computer attached to the scientific instrument and
collects data on local disks. Over these data the MCS server will
eventually perform some automatic analysis for data quick-look,
then will store the results and the raw data in the database.
The quick-look tool can be implemented in C++, or it can be
a previously written software; in the latter case you
only need to implement a simple interface between this tool and
MCS. A technician can then connect via a simple telnet terminal to the
MCS server and check if everything goes fine. This can be done using
mnemonic command codes, very similar to the commands issued in an
ordinary Unix shell. These command codes can of course be customized. A
researcher can use the ``Client'' class to develop a client tool to
connect to the MCS server and perform a remote control of the
instrument or retrieve scientific data in a variety of formats (typically
formats are raw, ASCII, FITS or VOTable). As already mentioned,
if the researcher doesn't want to
use C++ to implement the client tool, he can use one of the available
interfaces to MCS from other programming languages such as C, IDL,
Fortran or PHP. One of the main feature of these interfaces is that
the function names are the same in all languages. Finally there can be
a web server that connects to MCS through a PHP interface and
provides bookkeeping and outreach information as web pages to external users.
Each afore-mentioned user can of course have different privileges, and so
being able to view and retrieve different kinds of data.

\section{MCS and the Virtual Observatory}
MCS doesn't pretend to be an all-comprehensive, multi-wavelength
repository for astronomical data like the Virtual Observatory is, nor
a general purpose data analysis software. In this sense MCS is (of
course) not a Virtual Observatory competitor. Instead it was implemented
to easily setup real time information services, while reusing much
of the already existing software tools and database tables.
However in the near future MCS will implement the
PLASTIC\footnote{plastic.sourceforge.net} interface to connect and use
the Virtual Observatory facilities. 

MCS can also be seen as a starting point to implement or
simply make practice with a distributed computing environment like
in the ``GRID'' based projects. A network of computers, each running an
MCS server, can connect to all the others as client to perform
computation or database queries. Of course, also in this case MCS and ``GRID''
are not competitors because they are supposed to be used for different
purposes: in fact implementing a service and setting up an MCS server is
much more simple than setting up a node or putting a service on the ``GRID''.
On the other hand the ``GRID'' projects offer many more features than
MCS does.

\section{The MCS companion tools}
We have developed a number of software tools that can be used with MCS to
improve its functionalities (of course these software can also be used
as standalone):
\begin{itemize}
\item \textbf{MyRO} (My Record Oriented privilege system): this
  software provides a natural extension to the MySQL privilege system,
  offering the possibility to specify privileges on a record level. It
  supports users and groups like those of a Unix system, giving the
  possibility to set a read and/or write permission to each record.
  Needs MySQL version 5.

\item \textbf{DBEngine for FITS and VOTable}: a DBEngine is a software
  library linked into the database server (MySQL in this case) which
  let users read and write table in a format different from the
  MySQL proprietary. Our software let you read and write FITS and
  VOTable files as if they were database tables (this software is still
  under development).

\item \textbf{Database interface to HEALPix and HTM}: this software
  provides some of the functions of the HEALPix and HTM library as MySQL
  functions, to be used directly in SQL queries. Functions allowing
  (fast) circular and rectangular selections of entries in HTM indexed
  database tables will also be available soon.
\end{itemize}

\section{The future}
MCS is in continuous development. It will soon implement interfaces for
other languages such as Python, Perl, Java, Tcl/Tk. Another feature
that will be implemented soon is the integration with MyRO to handle a more
complex privilege system.

User contributed libraries for the various supported languages are being
built. This will allow an even easier access to the MCS functionalities
to non expert programmers. Moreover, commonly used,
independently developed external packages will also be included and
made accessible trough MCS. They include the afore mentioned HEALPix
and HTM libraries for sky pixelization scheme, the
Naval Observatory Vector Astrometry Subroutines (NOVAS --
 aa.usno.navy.mil/software/novas) used for astrometric
calculations and transformations, the World Coordinate System
 (WCS -- fits.gsfc.nasa.gov/fits\_wcs.html) library and tools.

\section{MCS is already used by...}
MCS is already used for managing the data produced by the ROSS camera.
The SPOrt experiment, whenever it will be resumed, will have the
data fully managed by an MCS based system.
The Astrometry group of the Turin Astronomical
Observatory is making use of the MCS client library through the IDL
interface to cross match HTM indexed catalogues. Other astronomical
experiments from the gamma-rays to the radio band have shown interest
to use MCS for their data management. One of them is the Italian
gamma-rays satellite AGILE.

\section{Conclusions}
Several astronomical projects are still supported by software
developed under an old conception: handling all data files with
``ad-hoc'' developed scripts, running stand alone applications which
can read only proprietary ASCII or binary files, relying on users
responsibility
to follow the format and naming specifications for files, absence of
any mechanism to automatically update a web site, etc. However today
information technology offers a lot of new tools which can give
several benefits to astronomers: database systems to efficiently store
and retrieve huge quantities of data, libraries to easily distribute
data across the network, widely used standard formats for information
interchange using freely available libraries, new high level languages to
quickly implement sophisticated data handling and processing, and so on.
All of these
tools are being effectively used by the Virtual Observatory and GRID
projects. They will become the standard ``de-facto'' in the near
future, so software developers involved in astronomical projects should
start to get used to these new technologies. MCS aims at being the simplest
and fastest tool to let developers build new software to support
scientific projects using these new technologies.

\begin{acknowledgements}
The SPOrt experiment was supported by ASI. The REM Observatory is supported
by INAF.
\end{acknowledgements}
